# Seasonal Variations in Relative Weight of Lake Trout (*Salvelinus namaycush*), Kokanee Salmon (*Oncorhynchus nerka*), Rainbow Trout (*Onocorhynchus mykiss*), and Brown Trout (*Salmo trutta*) in Blue Mesa Reservoir, Colorado

Madeline Midas, Asia Williams, Cindy Cooper
U. S. Air Force Academy, 2354 Fairchild Drive, USAF Academy, CO, 80840

Michael Courtney
BTG Research, PO Box 62541 Colorado Springs, CO 80920
Michael_Courtney@alum.mit.edu

## ABSTRACT

Blue Mesa Reservoir is the largest body of water in Colorado and is located on the western slope of the Rocky Mountains at an elevation of 7520 feet. Blue Mesa Reservoir contains recreationally important populations of lake trout (*Salvelinus namaycush*), kokanee salmon (*Oncorhynchus nerka*), rainbow trout (*Onocorhynchus mykiss*), and brown trout (*Salmo trutta*). A management challenge in recent years has been the overpopulation of lake trout, which has led to a steep decline in abundance of kokanee salmon (a preferred food source) followed by a steep decline in body condition of the lake trout. Relative weight, Wr, is commonly used to assess body condition and prey availability of fish in a variety of ecosystems, but its seasonal variability has not been well documented in cold water species. Relative weight is 100 times the actual weight of a fish divided by expected weight (or standard weight) of that species at the measured total length. The present study uses multi-year, multi-season length and weight data provided by the Colorado Department of Parks and Wildlife to quantify seasonal variations in mean relative weight in these four species. It is found that lake trout consistently lose significant body condition between the spring and the fall of a given year, with a mean loss in relative weight of 11.7 +/- 2.4. In contrast, kokanee salmon tend to gain significant body condition between spring and fall with a mean gain in relative weight of 5.2 +/- 2.9. Rainbow trout tend to lose body condition between summer and fall with a mean loss of relative weight of 7.4 +/- 1.7. Brown trout tend to gain body condition between spring and fall with a mean gain of 5.6 +/- 1.8. These findings can be interpreted in terms of the feeding habits, forage availability, and metabolic activity of each species.

## INTRODUCTION

Blue Mesa Reservoir is located about 30 miles west of Gunnison, CO and is a major attraction for anglers in the region. The reservoir is at an elevation of 7,520 feet, which results in a high production of zooplankton (Colorado Parks and Wildlife, 2012). This zooplankton is important to the food web, because it is the primary forage the kokanee salmon. The four species that were considered in the present study were the kokanee salmon (*Oncorhynchus nerka*), brown trout (*Salmo trutta*), mackinaw (*Salvelinus namaycush*), and rainbow trout (*Onocorhynchus mykiss*).

Kokanee salmon were introduced to Blue Mesa Reservoir in 1965, and a top goal of Colorado Parks and Wildlife has been to keep their population balanced (Colorado Parks and Wildlife, 2012). Kokanee are the primary target of recreational anglers who travel to Gunnison County and bring in annual revenue of about $8 million dollars to the local economy (Colorado Parks and Wildlife, 2012).

The introduction of the lake trout has become detrimental to the kokanee population in Blue Mesa Reservoir. Colorado Parks and Wildlife has decided not to completely eliminate lake trout, because significant numbers of anglers are interested in catching the trophy sized lake trout (up to 50 pounds) and because the lake trout are well fed by the kokanee salmon, Blue Mesa Reservoir has gained a reputation as a trophy lake trout fishery. However, Colorado Parks and Wildlife has taken steps to reduce the number of lake trout with spring and fall netting operations that cull lake trout less than 30 inches in length. In this way, anglers are still able to fish for trophy mackinaw, the mackinaw have less competition thus increasing their ability to grow so large, and the kokanee have fewer predators (Colorado Parks and Wildlife, 2012).

Relative weight is 100 times the actual weight of a fish divided by the expected weight of a fish in good condition of a given total length. The expected weight is expressed as the standard weight which is a weight-length formula derived by consideration of a number of populations of a given species over its range (Anderson and Neumann, 1996) and is intended to represent a fish in good condition by estimating the weight of the 75$^{th}$ percentile fish of a given length. Thus, a fish with a relative weight of 100.0 is a fish in good condition, fish with relative weights above 100.0 are in very good to excellent condition, and fish with relative weights below 100.0 are on the average to thin side. As an inherent measure of fish plumpness, relative weight is commonly used to assess prey availability, evaluate management decisions, and compare with other populations (Blackwell et al., 2000).

It is generally expected that body condition in fish varies throughout the year and that comparisons between populations or from year to year in a given population requires samples to have been taken at about the same time of year.  However, seasonal variations in relative weight have not been widely documented for the cold water species studied here.  The purpose of the present study is to describe the seasonal variations in relative weight in lake trout, kokanee salmon, rainbow trout, and brown trout, and also to consider these seasonal variations by analyzing the data by length class when the quantity of data is sufficient for length class analysis.  Past studies (Johnson, 2005; Brauch, 2011; Brauch, 2012) have reported relative weights for different years; however, little or no consideration was given to season when comparing relative weight data from different years.

## METHOD

Survey data from 1994 to 2011 was obtained from the Colorado Division of Parks and Wildlife.  The Colorado Division of Parks and Wildlife (CDPW) has employed a variety of techniques over the years to acquire weight-length data, including gill nets, trammel nets, entrainment nets, and electrofishing.  Different sampling techniques are known to be selective for given length ranges; however, for a given length range, sampling techniques are not selective for plumpness, so that relative weight determinations are not believed to be biased by variations in sampling technique.

Data acquired from CPDW was sorted by species, date, and whether both length and weight data were available (which is required to compute relative weight).  Where needed to compute relative weights by length class, data was further sorted by length class (Gabelhouse, 1984).  The relative weight was computed for each fish as the actual weight divided by the standard weight times 100   (Anderson and Neumann, 1996).  Neither a standard weight equation nor length class definitions for brown trout in lentic systems had been published in Anderson and Neumann (1996), these important standards were taken from Hyatt and Hubert (2001).  Mean relative weights are the arithmetic mean of the given sample (species, year, season, and possibly length class).  Reported uncertainties are the standard error of the mean.  Data is reported here only for years and species for which sufficient data was available to compute relative weights for different seasons.  There were many cases where there was only sufficient data to compute relative weights of a given species for a single season of a year.

## RESULTS

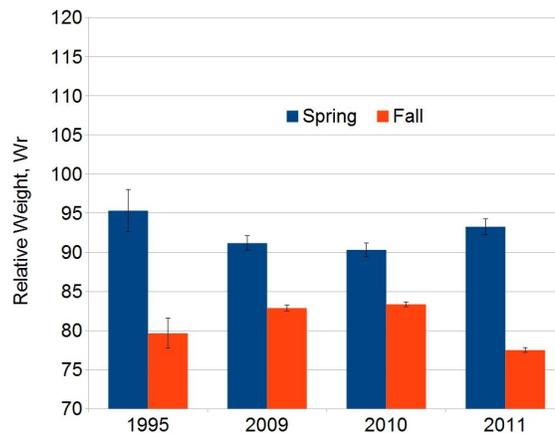

*Figure 1: Mean relative weight of lake trout (all length classes) in Blue Mesa Reservoir by season and year.  Error bars are the standard error of the mean.  Note significant decreases in relative weight from spring to fall every year.*

**Lake Trout**
Length and weight data were available for 4321 lake trout in 1995, 2009, 2010, and 2011 permitting detailed analysis of relative weights by length class and season.  Figure 1 shows the mean relative weight of all length classes of lake trout by season and year.  The relative weight shows significant decreases from spring to fall in each of the four years.  Lake trout tend to enter spring in relatively good body condition (mean $90.3 < Wr < 95.3$), but decline significantly in body condition through the summer and into the fall (mean $77.5 < Wr < 83.4$).  This may be due to a combination of possible factors: 1)  Their preferred prey (kokanee salmon) are more active in the warmer months and better able to avoid predation. 2) Age zero kokanee salmon are not yet large enough during the summer to provide adequate forage during the summer and the age one kokanee salmon have grown too large for all but the largest lake trout. 3) The metabolic needs of lake trout are higher in the warmer months, especially as they need to pursue kokanee and other forage species into the upper, warmer waters. 4) The population of lake trout is overpopulated in Blue Mesa Reservoir and there is inadequate forage to maintain body condition in the warmer months when metabolic needs are higher.

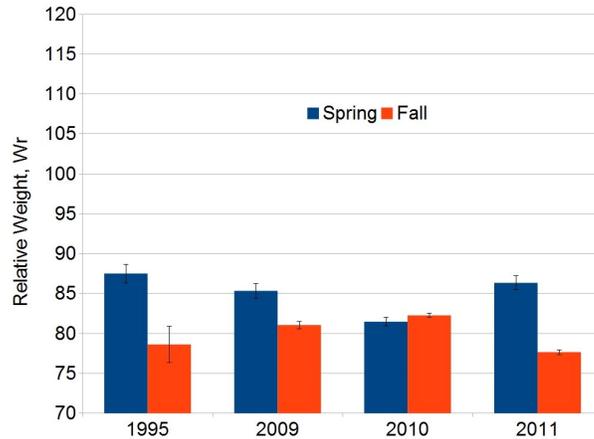

*Figure 2: Mean relative weight of lake trout (stocker length class, 300-500 mm total length) in Blue Mesa Reservoir by season and year. Error bars are the standard error of the mean.*

When available data permits, it is appropriate to analyze relative weights by length class, as described in Anderson and Neumann (1996). The stocker length class in lake trout includes total lengths from 300-500 mm, and the resulting length class analysis is shown in Figure 2. The mean spring relative weights of the stocker length class ($81.5 < Wr < 87.5$) are generally smaller than lake trout as a whole, suggesting that the stocker length class is not competitive with the longer length classes for forage. Figure 2 also shows a decrease in relative weight between spring and fall for all years except for 2010 with the fall mean relative weights averaging from $77.6 < Wr < 82.3$. The decrease in relative weight between spring and fall averages 5.3 +/- 2.3. The fact that Wr did not drop between spring and fall of 2010 may be due to some combination of various factors: 1) There is some relative weight below which lack of body condition will lead to increased mortality. It is possible that rather than become increasingly thin, fish in this length class may have died and been removed from the population, therefore unable to reduce the mean relative condition factor below the spring value of 81.5. Starvation related morality would have the effect of reducing pressure on the forage for the surviving fish in this length class. 2) Perhaps there was a relative abundance in small fish that make up the preferred forage sources of this length class of lake trout. Brown trout, which probably share common food sources with lake trout in this length class, saw their largest increase in relative weight from summer to fall 2010, from 68.3 in spring to 76.0 in fall. Entering spring with a mean relative weight of 68.3 probably suggests some possibility of starvation related mortality in the brown trout also. 3) It is also possible that the starving brown trout under 200 mm total length actually became an important food source for the lake trout.

The quality length class in lake trout includes total lengths from 500-650 mm, and the resulting length class analysis is shown in Figure 3. The mean spring relative weights of the quality length class ($84.5 < Wr < 89.6$) were slightly higher than the stocker length class, suggesting a slight competitive advantage for available forage. Figure 3 also shows a decrease in relative weight between spring and fall for all years with decreases in relative weight ranging from 3.2 to 14.9 and averaging 7.5 +/- 3.7. Body condition of the quality length class of lake trout in Blue Mesa Reservoir seems to decline between the spring and the fall, and then increase between the fall and the subsequent spring. The increase during the cooler months may be due to a combination of reduced metabolic demands and increased availability of preferred forage sources.

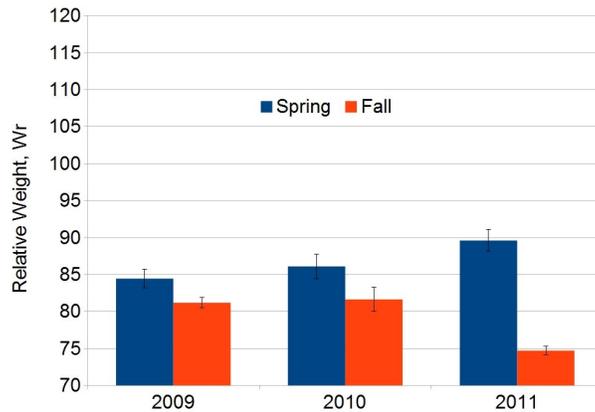

*Figure 3: Mean relative weight of lake trout (quality length class, 500-650 mm total length) in Blue Mesa Reservoir by season and year. Error bars are the standard error of the mean.*

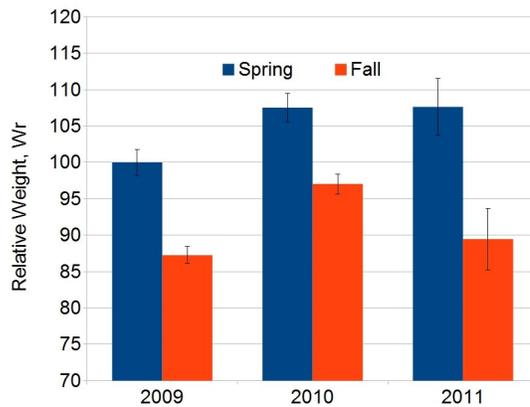

*Figure 4: Mean relative weight of lake trout (preferred length class, 650-800 mm total length) in Blue Mesa Reservoir by season and year. Error bars are the standard error of the mean.*

The preferred length class in lake trout includes total lengths from 650-800 mm, and the resulting length class analysis is shown in Figure 4. The mean spring relative weights of the preferred length class ($100.0 < W_r < 107.6$) are significantly higher than the stocker and quality length classes, suggesting a competitive advantage for available forage. This is the smallest length class of lake trout which are consistently entering the spring in good body condition, suggesting that the ability to successfully consume larger kokanee salmon is important. Figure 4 also shows a decrease in relative weight between spring and fall for all years, with decreases in relative weight ranging from 10.5 to 18.2 and averaging 13.8 +/- 2.3. These larger lake trout not only come through the winter with higher body condition than smaller length classes, they lose more of their body condition over the warmer months. The decrease during the warmer months may be due to a combination of higher metabolic demands and reduced availability of preferred forage sources. It is known that these larger lake trout prefer the deeper, cooler waters on the western side of the reservoir (closer to the dam) during the summer months. In contrast, the shorter length classes are more commonly found spread throughout the reservoir.

      The memorable and trophy length classes combined in lake trout include total lengths greater than 800 mm. The resulting length class analysis is shown in Figure 5. The mean spring relative weights of these length classes ($104.1 < W_r < 114.5$) are significantly higher than the stocker and quality length classes and higher than the preferred length class, suggesting a competitive advantage for available forage. Figure 5 also shows a decrease in

relative weight between spring and fall for all years, with decreases in relative weight ranging from 8.3 to 23.3 and averaging 15.8 +/- 4.3.  These largest lake trout both have higher body condition than smaller length classes in spring, and they lose more of their body condition over the warmer months.  The decrease during the warmer months may be due to a combination of higher metabolic demands and reduced availability of preferred forage sources.

    The observation that the length classes that enter spring with higher relative weights also lose more relative weight between spring and fall is suggestive of a possible trend that the loss of relative weight depends on the amount of relative weight available to lose.  To investigate this possibility, the loss of relative weight between spring and fall is plotted for all length classes in Figure 6, along with a best fit trend line.  There is a significant relationship ($r = 0.810$, $p < 0.05$) between spring relative weight and the loss of relative weight from spring to fall, but it is not clear whether this relationship is caused by 1) greater metabolic demands of plumper fish over the warmer months 2) greater metabolic demands of longer fish which tend to be plumper 3) difficulty in longer fish finding sufficient forage over the summer months 4) greater mortality of fish below a certain relative weight (the horizontal intercept of the trend line in Figure 6 suggests it is increasingly difficult to continue to lose relative weight below 74) or 5) rapid length growth of plumper fish so that plump fish entering spring have a growth spurt over the summer and are thinner entering the fall.

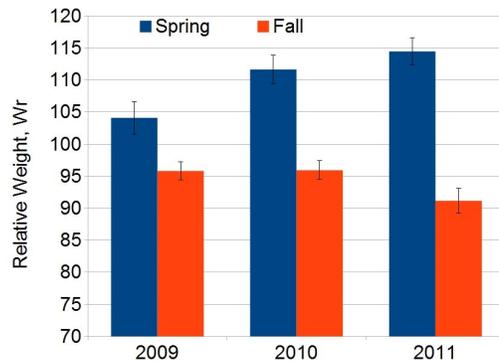

*Figure 5: Mean relative weight of lake trout (memorable and trophy length classes combined, total length > 800 mm) in Blue Mesa Reservoir by season and year.  Error bars are the standard error of the mean.*

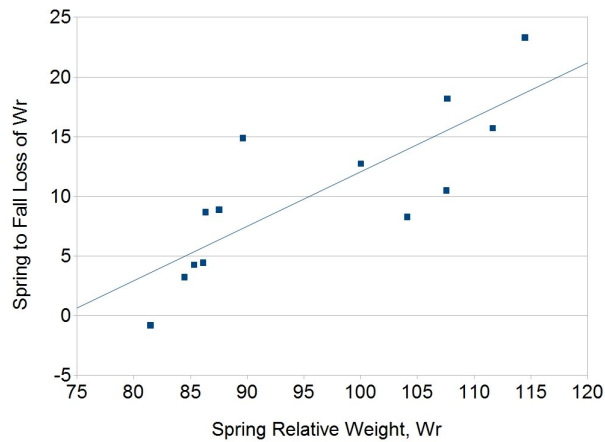

*Figure 6: Spring to fall loss of relative weight vs. spring relative weight for all length classes and years in lake trout in Blue Mesa Reservoir.*

**Kokanee Salmon**

Length and weight data were available for 750 kokanee salmon in 1994, 1995, 2001, and 2011 permitting analysis of relative weights by season and year. Figure 7 shows the mean relative weight of all length classes of kokanee by season and year. Except for 1994, there is an increasing trend from spring to fall each year, with kokanee entering spring each year with fair to good relative weights (mean $92.1 < W_r < 99.6$), then increasing by an average of 5.2 +/- 2.9 between spring and fall. This may be due to a combination of possible factors: 1) Blue Mesa Reservoir offers excellent conditions for the production of daphnia, their preferred food source (Johnson and Koski, 2005). 2) A combination of predation by lake trout and harvest by anglers keeps the kokanee salmon population at sustainable levels so that they do not overly stress their food sources. 3) The limnology of the reservoir is conducive to maintaining a healthy and plump population of kokanee salmon.

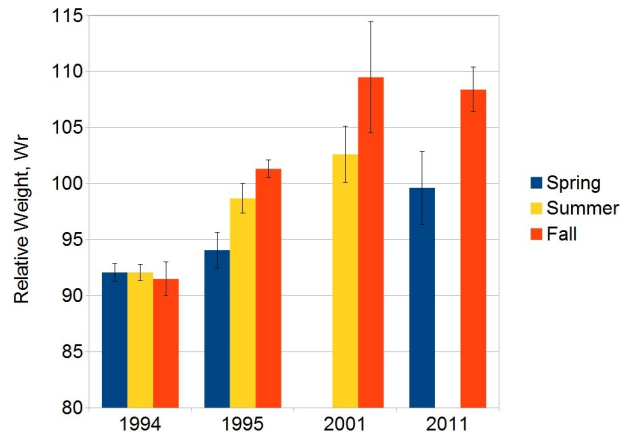

*Figure 7: Mean relative weight of kokanee salmon (all lengths) in Blue Mesa Reservoir by season and year. Error bars are the standard error of the mean.*

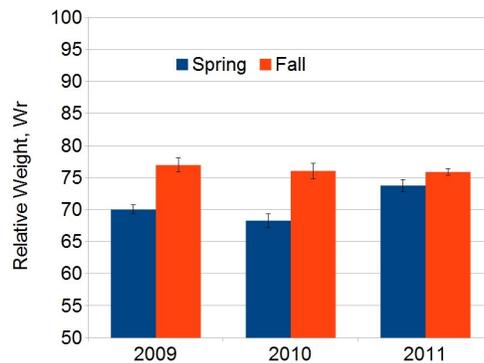

*Figure 8: Mean relative weight of brown trout (all lengths) in Blue Mesa Reservoir by season and year. Error bars show the standard error of the mean.*

**Brown Trout**

Length and weight data were available for 1090 brown trout for the spring and fall of 2009, 2010, and 2011 permitting analysis of relative weights by season, length class and year. The relative weight equation for brown trout was taken from Hyatt and Hubert (2001). Figure 8 shows the mean relative weight of all length classes of by season and year. Brown trout are very thin in Blue Mesa Reservoir in both the spring and the fall of all sampling years. Other studies have also found brown trout to be very thin in Blue Mesa (Johnson and Koski, 2005; Abbott et al., 2013) Brown trout reproduction in the reservoir is self-sustaining and there is neither ample natural predation nor human harvest to maintain brown trout populations at a level where the available forage is capable of maintaining good relative weights. Consequently, brown trout in the reservoir tend to be very thin. The lower tail of the histograms of relative weights shown in Figure 9 probably suggest the survivability of low relative weights for brown trout under reservoir conditions. Survivability becomes increasingly less likely for Wr < 60. It is possible that selection pressure over time has yielded brown trout population genetics in the reservoir that are more capable of surviving and successfully reproducing at lower relative weights. It should be pointed out that the survivability issue means that distributions of relative weights in most populations are unlikely to be normally distributed (Gaussian) or even symmetric. The fact that some level of thinness implies greatly reduced survival chances provides a much harder lower limit. In contrast, increasing plumpness does not significantly reduce survival chances, and the relative weight distribution can easily have a much longer tail at high relative weights than it can at low relative weights.

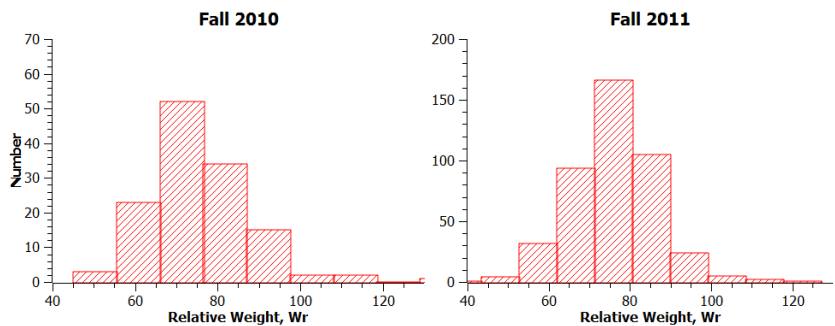

*Figure 9: Histogram of relative weights for fall 2010 and 2011.*

There is a notable increasing trend from spring to fall each year, with brown trout relative weight increasing by an average of 5.6 +/- 1.8 between spring and fall. Brown trout are highly piscivorous and are likely preying heavily on the young of the year of all the available species in the reservoir having substantial habitat overlap with them. This food source is abundant in the spring and summer, but declines in availability, and many of the available young of the year grow to lengths that put them out of reach of most brown trout by the fall.

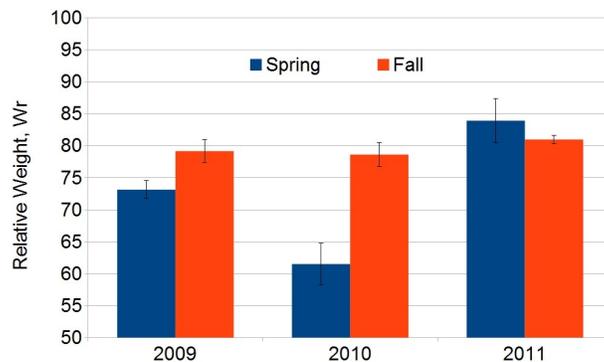

*Figure 10: Mean relative weight of brown trout (quality length class, 300-400 mm total length) in Blue Mesa Reservoir by season and year. Error bars are the standard error of the mean.*

Seasonal mean relative weights for the quality length class of brown trout (300 – 400 mm in total length) are shown in Figure 10. Mean relative weights show a significant increase from spring to fall in 2009 and 2010, followed by a slight decrease in 2011. Low rainfall in the Gunnison River watershed contributed to a significant drop in water level from spring to fall 2011, and a decrease in nutrient flux into the reservoir may have limited primary productivity. Over the same time interval (spring to fall 2011), all length classes of lake trout also saw larger than average decreases in relative weight suggesting that the common factor is most likely a decrease in prey availability for these piscivorous species. Another factor in failing to increase in mean relative weight from spring to fall 2011 may be that the quality length class of brown trout entered the spring with a relatively high mean relative weight (83.9 +/- 0.6) in spring 2011. In a similar same way that lake trout which enter the spring plump tend to lose more body condition over the summer because they have more body condition to lose, it may be that brown trout that enter the spring relatively plump have a harder time making further gains in body condition.

Mean relative weights for the preferred length class of brown trout (400 – 500 mm in total length) are shown in Figure 11 by season and year. As in the case of the quality length class, mean relative weights show significant increases from spring to fall in 2009 and 2010, followed by a slight decrease in 2011, probably for the same reasons.

Figure 12 shows mean relative weights for the combined memorable and trophy classes of brown trout (total lengths > 500 mm) in Blue Mesa Reservoir by season and year. Note the steep increase in body condition from spring to fall 2010, followed by a precipitous decline the following spring (2011) and then an additional decline from spring to fall 2011. The steep increase from spring to fall 2010 is probably due to uncommonly abundant forage. The piscivorous lake trout from 300 - 500 mm in total length also saw an increase in mean Wr over the same interval (see Figure 2), even though lake trout actually decline in body condition from spring to fall for most length classes and years. However, brown trout in the memorable and trophy length classes were unable to maintain the uncommonly high level of average body condition achieved in fall 2010, declining to 75.7 +/- 3.5 in spring 2011 and then to 62.5 in fall 2011.

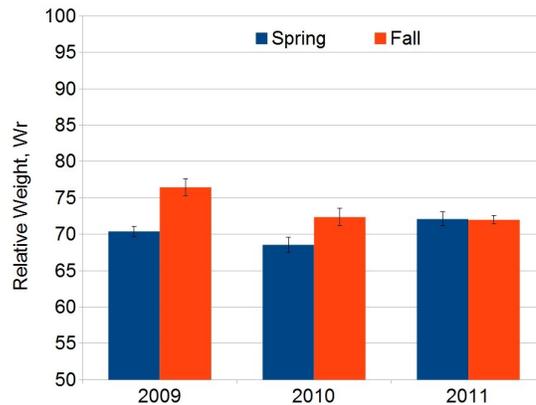

*Figure 11: Mean relative weight of brown trout (preferred length class, 400-500 mm total length) in Blue Mesa Reservoir by season and year. Error bars are the standard error of the mean.*

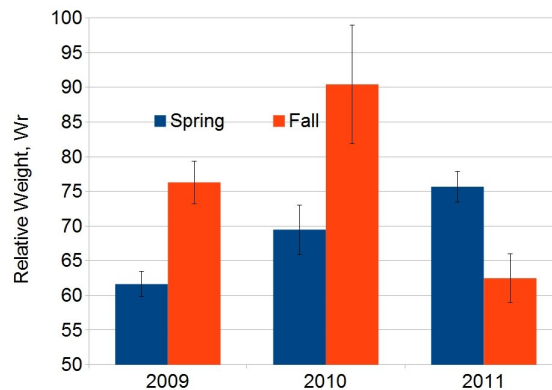

*Figure 12: Mean relative weight of brown trout (memorable and trophy length classes combined, total lengths > 500 mm) in Blue Mesa Reservoir by season and year. Error bars are the standard error of the mean.*

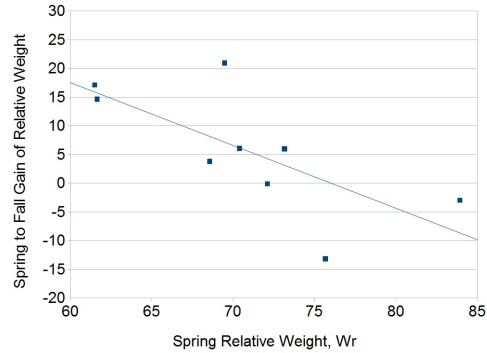

*Figure 13: Spring to fall gain of relative weight vs. spring relative weight for all length classes and years of brown trout in Blue Mesa Reservoir.*

The observation that the length classes that enter spring with higher relative weights show smaller gains in relative weight between spring and fall is suggestive of a possible trend that the gain of relative weight depends on the amount of relative weight available to gain. In other words, it seems that thin fish can gain body condition more readily than plump fish. To investigate this possibility, the gain of relative weight between spring and fall is plotted for all length classes in Figure 13, along with a best fit trend line. There is a significant relationship ($r = 0.707$, $p < 0.05$) between spring relative weight and the gain of relative weight from spring to fall, but it is not clear whether this relationship is caused by 1) greater metabolic demands of plumper fish over the warmer months 2) rapid length growth of plumper fish so that plump fish entering spring have a growth spurt over the summer and are thinner entering the fall.

**Rainbow Trout**
Figure 14 shows the mean relative weight of rainbow trout in the reservoir by season and year. Length classes are combined. Unfortunately, availability of survey data is inconsistent with either spring or summer data available in various years. There is a trend toward decreasing relative weight from spring to fall and from summer to fall. This decrease is consistent with the findings of Courtney et al. (2013) who quantified the month to month changes in relative condition factor of rainbow trout, cutbow trout, and cutthroat trout in Elevenmile Reservoir, Colorado. The decrease in relative condition factor over the warmer months is probably dominated by higher metabolic demands in the warmer water and the tendency of fish to add length rather than body condition during these months (Parker et al. 2011) rather than having more food in the winter months and less food in the warmer months. Unlike lake trout and brown trout whose populations are sustained via natural reproduction, rainbow trout are incapable of sustaining reasonable populations via natural reproduction, and most of the rainbow trout in the reservoir are stocked from nearby hatcheries. Due to high predation losses to lake trout, most stocked rainbow trout are of catchable size (200-300 mm) at the time of stocking.

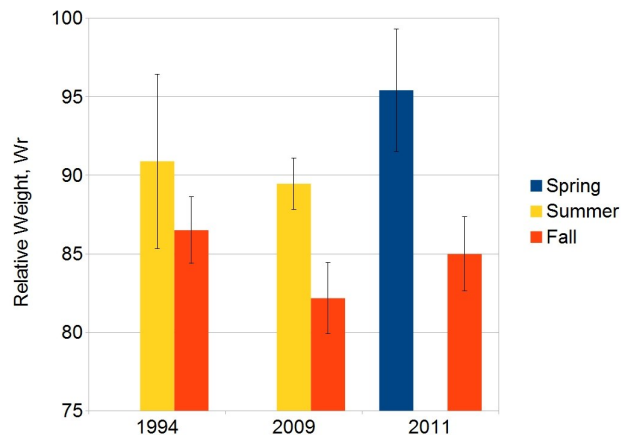

*Figure 14: Mean relative weight of rainbow trout (all length classes combined) in Blue Mesa Reservoir by season and year. Error bars are the standard error of the mean.*

# DISCUSSION

Lake trout and rainbow trout seem to lose body condition between spring and fall samplings; whereas, brown trout and kokanee salmon tend to gain body condition. This general trend can occasionally be upset by swings in forage availability due to other factors such as drought. The food web and population dynamics of the lake trout and the kokanee salmon have been extensively studied (Johnson and Koski 2005) due to the economic importance of these species. In contrast, food web and population dynamics of brown trout have not been extensively studied in the reservoir.

There appears to be some competitive interaction between the largest brown trout (total length > 500 mm) and the stocker class of lake trout (300 – 500 mm total length). Presumably, the larger gape width (relative to body size) of the lake trout has them competing for the same forage as the larger brown trout. However, until fish reach sufficient size to consume age 1 kokanee, there simply are not enough smaller forage fish available in the reservoir to support the existing populations of lake trout and brown trout in good body condition. Brown trout are most certainly seeing significant population losses to starvation, and lake trout are seeing body conditions much too small to support reasonable grown rates.

The liberal limits on small lake trout (unlimited lake trout < 27" long may be harvested) and the aggressive annual cullings (> 1000 fish annually in 2009, 2010, and 2011) of smaller lake trout are warranted by the limitations of available forage and are probably necessary to ensure the health of the food web. The limit on brown trout should probably be raised from 4 fish per angler per day (or eliminated completely), and the food web might benefit additionally from aggressive net based culling of brown trout also, especially if lake trout are viewed as being of greater recreational interest to anglers thus of greater economic importance to the region.

Resource managers would do well to set goals for improving the mean fall relative weights of brown trout and lake trout in the reservoir and manage the populations to reach these goals. Such goals would contribute to growth rates, improve angler satisfaction by producing fish that are not so overtly thin as to appear snakelike, and ensure that the forage base is not adversely affected. We suggest a mean fall relative weight of 85 would be a much healthier target for brown trout (all length classes combined) than the current status, and a mean fall relative weight of 90 would be a much healthier target for the stocker length class (300 – 500 mm total length) of lake trout.

## ACKNOWLEDGEMENTS

This work was funded by BTG Research (www.btgresearch.com) and the United States Air Force Academy. The authors are grateful to Harry Vermillion at the Colorado Division of Parks and Wildlife for providing survey data.